\documentclass[transmag]{IEEEtran}
\usepackage{latexsym}
\usepackage[T1]{fontenc}
\usepackage{amsfonts,amssymb,amsmath}
\usepackage{stfloats}
\usepackage{cite}
\usepackage{graphicx}
\usepackage{psfrag}
\usepackage{subfigure}
\usepackage{array}
\usepackage{algpseudocode}
\usepackage{setspace}
\usepackage{blindtext}
\usepackage{url}
\usepackage{verbatim}
\usepackage{bm}
\usepackage{multirow}
\usepackage{booktabs}
\usepackage{makecell}
\usepackage{ntheorem}
\usepackage{textcomp}
\usepackage{color}
\usepackage{epstopdf}
\def\BibTeX{{\rm B\kern-.05em{\sc i\kern-.025em b}\kern-.08em T\kern-.1667em\lower.7ex\hbox{E}\kern-.125emX}}
\begin{document}
\title{6G: Connecting Everything by 1000 Times Price Reduction}
\author{Shunqing~Zhang,~\IEEEmembership{Senior~Member,~IEEE,}
        Chenlu~Xiang,~\IEEEmembership{Student~Member,~IEEE,}
        and~Shugong~Xu,~\IEEEmembership{Fellow,~IEEE}
\thanks{Manuscript received February 15, 2020, accepted March 05, 2020. This work was supported in part by the National Natural Science Foundation of China (NSFC) Grants under No. 61701293 and No. 61871262, the National Science and Technology Major Project (Grant No. 2018ZX03001009), the National Key Research and Development Program of China (Grant No. 2017YFE0121400), the Huawei Innovation Research Program (HIRP), and research funds from Shanghai Institute for Advanced Communication and Data Science (SICS).}
\thanks{S. Zhang, C. Xiang, and S. Xu are with Shanghai Institute for Advanced Communication and Data Science, Key laboratory of Specialty Fiber Optics and Optical Access Networks, School of Information and Communication Engineering, Shanghai University, Shanghai, 200444, China (e-mail: \{shunqing, xcl, shugong\}@shu.edu.cn)}}
\maketitle
\begin{abstract}
The commercial deployment of 5G communication networks makes the industry and academia to seriously consider the possible solutions for the next generation. Although the conventional approach indicates that 6G vision and requirements can be figured out by simply multiplying a certain amount of magnitude on top of the previous generations, we argue in this article that 1000 times price reduction from the customer's view point is the key to success. Guided by this vision, we categorize the current candidate technologies in a well organized manner and select AI-assisted intelligent communications as an example to elaborate the drive-force behind. Although it is impossible to identify every detail of 6G during the current time frame, we believe this article will help to eliminate the technical uncertainties and aggregate the efforts towards key breakthrough innovations for 6G.
\end{abstract}


\section{Introduction} \label{sect:intro}
\IEEEPARstart{W}{ith} the commercial deployment of the fifth generation (5G) wireless communication networks around the world, both the industry and academia begin to think about what the sixth generation (6G) wireless communication network should be and what are the service requirements behind. If we recall the standardization processes of 5G networks, three typical scenarios have been identified in the initial stage, including enhanced mobile broadband (eMBB), ultra-reliable and low-latency communications (URLLC), and massive machine-type communications (mMTC), which issue the design guideline of 5G technologies. Although the scenario identifications have not been finalized for 6G networks, several pioneering works, such as \cite{David2018,Saad2019,chen2019learning}, forecast the vision to connect everything worldwide via nearly instantaneous, reliable, and unlimited wireless resources.

To fulfill the vision of connecting everything worldwide, 6G will be responsible for the extreme communication requirements that can be rarely supported by 5G, such as real-time digital twins for smart manufacturing \cite{ahmed2016internet}, seamless holographic projection for smart working \cite{wakunami2016projection}, or wireless brain-computer interactions for smart living \cite{Saad2019}. In order to nail down the 6G  vision with more specific performance metrics, we provide the following list by comparing with the conventional 5G requirements.
\begin{itemize}
    \item{\em Throughput} A peak throughput of 1 Tb/s will be necessary for 6G, which is roughly 1000 times faster than 5G, while for user-experienced throughput, 100 times improvement, e.g. from 100 Mb/s to 10 Gb/s, will be more practical.
    \item{\em Latency} From the end-to-end point of view, the communication latency will be reduced by at least 10 times, which corresponds to 1 ms. To ensure this target, the link level transmission latency needs to be reduced from 1 ms to less than 50 ns.
    \item{\em Reliability} Reliability of 99.99999\% will be guaranteed to support most of unmanned systems, including autonomous driving vehicles and collaborative robotics.
    \item{\em Spectrum and Energy Efficiency} To achieve another 10 times spectrum and energy efficiency improvement for point-to-point links will be extremely challenging, however, 10 times area spectrum efficiency and 100 times area energy efficiency will be possible.
    \item{\em Connection Density} The scope of ``connections'' will be extended to cover different types of network nodes and support vehicle-to-vehicle or other direct communications with the peak density increased by another 1000 times to reach $10^7 \sim 10^8$ links/km$^2$.
    \item{\em Mobility} The maximum supported speed will be further enhanced from 500 km/h (ultra high-speed train) to subsonic speed, e.g. 1000 km/h (plane).
\end{itemize}

Together with ubiquitous coverage requirements, the above diversified metrics make 6G networks a disruptive technology shift to utilize more spatial dimensions, more frequency bands, and more flexible frame structures. Several representative technologies have been widely discussed to fulfill this target. For example, {\em Space-Air-Ground integrated network} \cite{Zhang2017} is proposed to extend the spatial degrees of freedom by incorporating terrestrial, airborne, and satellite networks, which stretches conventional 2D networks into 3D space for more efficient and reliable connections \cite{zhang2020first}. Aiming to explore extra under-utilized high-frequency bands, {\em Terahertz (THz) and visible light communication (VLC)} are shown to be promising for 1 Tb/s throughput with ultra-scale multiple antennas and over tens of gigahertz (GHz) available bandwidth \cite{Akyildiz2018, Yang2019}. Meanwhile, intelligent control and utilization of massive time-frequency-space resources rely on a fundamental reconfiguration of wireless network architectures and {\em artificial intelligence (AI) driven communication} will be the possible answer \cite{Letaief2019}.

Instead of providing a straight-forward literature overview for all the above technical directions, we re-investigate the vision of 6G communications from economic point of view in this report. Specifically, by examining the previous generations of wireless communications, we predict that 6G networks will offer a wireless connection with less than 0.1 US dollars/year, which is {\bf 1000 times} price reduction if compared with conventional 5G systems. Guided by this vision, we re-organize the current technologies in a more structured way and discuss the corresponding breakdown for 1000 times' goal. In addition, we select the most important technology, e.g., AI-assisted intelligent communications, to analyse the unique challenges for 6G networks, and offer a deep-dive using case studies. Although it is impossible to identify every detail of 6G during the current time frame, we believe this article will help to eliminate the technical uncertainties and aggregate the efforts towards key breakthrough innovations for 6G.

The rest of this article is structured as follows. Section~\ref{sect:1000} provides the vision of 6G communications to achieve 1000 times price reduction for connecting the societies, and Section~\ref{sect:AIIC} discusses the main drive-force to adopt AI-assisted intelligent communications. In Section~\ref{sect:CE}, we rely on two cases to demonstrate the possibly application of AI-assisted communication technologies, and discuss the potential challenges in Section~\ref{sect:cha}. Concluding remarks are summarized in Section~\ref{sect:conc}.

\section{Why 1000 Times?} \label{sect:1000}
In the past few decades, the rapid development of wireless technologies has made continuous evolution of wireless communication networks. Looking back on the previous generations, the influences of ``even-number'' generations wireless networks are in general better than ``odd-number'' generations. Is there any reason behind? What can we learn from history to formulate 6G vision? In this section, we provide a historic overview to answer the above questions and discuss the possible technology breakdown in what follows.

\subsection{Historical Overview}
From the viewpoint of non-technical telecommunication customers, the most intuitive experience for different generations of wireless networks is the monthly bills paid and the corresponding services delivered, rather than the advanced technologies used. Thus, we focus on the killing applications and the corresponding service charges to reflect their differences.

\subsubsection{1G \& 2G - 10 Times Reduction} The basic service for 1G and 2G networks is {\em voice calling}. Although significant technical efforts on top of 1G have been contributed to make 2G a fully digitized system, {\bf 10 times price reduction}, e.g. from 0.1 US dollars/minute to 0.01 US dollars/minute according to China Mobile's annual report \cite{cm2017report}, is the most valuable result for non-technical customers using voice calling. With the above achievement, the fraction of the world's population using such services was increased from well below 10\% (for 1G) to more than 50\% (for 2G) within 20 years \cite{gong2017socialist}.

\subsubsection{3G \& 4G - 1000 Times Reduction} {\em Data transmission} is often recognized as a key service delivered by 3G and 4G systems, together with the technical transition from circuit-switched to packet-switched networks. Despite the significant technical progresses from 3G to 4G, including multiple-input multiple-output (MIMO) and orthogonal frequency division multiplexing (OFDM), {\bf 1000 times price reduction}, e.g. from 0.1 US dollars/megabits to 0.1 US dollars/gigabits \cite{cm2017report}, is the most user-sensitive achievement. Similar to what happened in 1G and 2G, the initial 3G users is limited to businessmen for accessing e-mails and company resources, while the explosive growth of users occurs only after 4G networks are widely deployed.

\subsubsection{5G \& 6G - 1000 Times Reduction} A common view for 5G and beyond is that it will facilitate human-to-machine or even machine-to-machine communications for ``{\em connected society}'' rather than ``connected people''. Although the current deployment of 5G is still based on eMBB scenario with the similar pricing strategy as 4G networks, it will be more reasonable to charge based on connections rather than voice or data traffics. Referring to fibre to the home (FTTH) systems, the annual price for each terminal is around 100 to 200 US dollars in China \cite{ftth2019report}, which may become a suitable charging baseline for 5G with 100 US dollars/year per connection. By the year 2030 when 6G is expected to be deployed, more than 100 trillion sensors will be manufactured and connected to Internet \cite{David2018}. Hence,  {\bf 1000 times price reduction}, e.g. to reach 0.1 US dollars/year per connection for 6G, will be necessary to maintain a sustainable development of the smart society.

\begin{figure}
\centering
\includegraphics[width = 3.4 in]{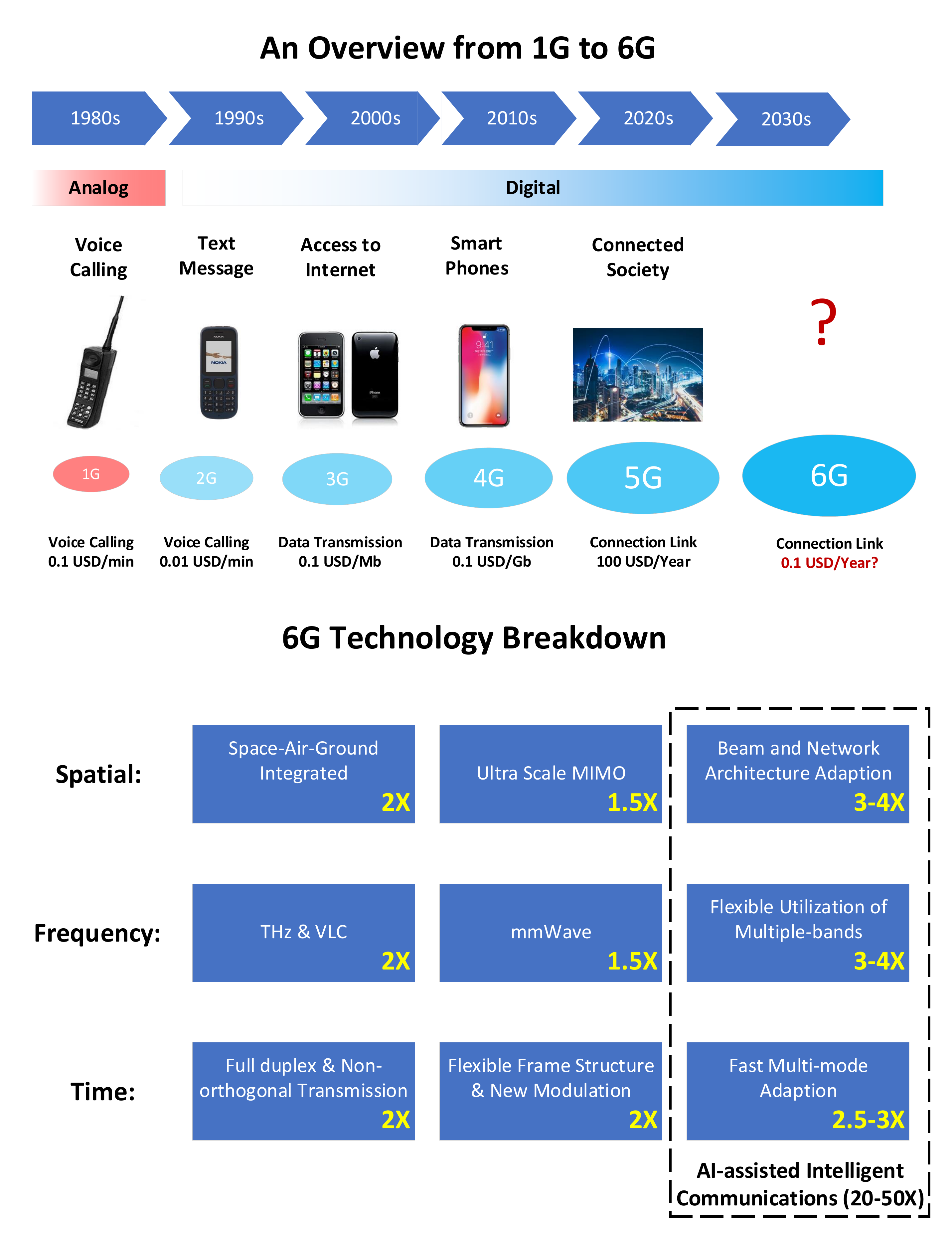}
\caption{An overview of typical devices and service prices for 1G to 6G communication networks and the corresponding technology breakdown for 6G.}
\label{fig:6g}
\end{figure}

\subsection{Technology Breakdown}
Figure~\ref{fig:6g} summarizes typical devices and service prices for each generation wireless network. Although it is quite challenging to achieve 1000 times price reduction for 6G networks, we break it down according to spatial, frequency, and time domains as elaborated below.
\begin{itemize}
    \item {\em Spatial - 10 Times} The scope of the Space-Air-Ground integrated network covers a wide range of flying sites and terminals (such as drone), satellite communications, which can offer two times reduction by requiring much less number of serving base stations. Ultra scale MIMO is expected to have 50\% throughput improvement without additional costs, thereby reducing the overall price by 1.5 times. Another three to four times reduction comes from the intelligent adaptation of beam and different network architectures, which eventually brings 10 times reduction.
    \item {\em Frequency - 10 Times} The cost reduction in frequency domain relies on the exploration of low cost spectrum. Although THz, VLC, and mmWave are able to provide sufficient bandwidth for wireless transmission, the most suitable case will be the indoor users with pedestrian mobility, which is equivalent to 70\% of the total wireless traffics \cite{lin2016energy}. Thus, three time reduction\footnote{We assume all the indoor traffics can be handled by higher frequency bands using WiFi-like techniques with marginal costs.} will be achievable using the higher frequency bands. In addition, by flexible utilization of multiple frequency bands, another three to four times reduction will be possible.
    \item {\em Time - 10 Times} Given a fixed time-frequency resource, the achievable throughput using current technologies are close to the Shannon's limit for point-to-point case. Hence, one desired solution is to build more point-to-point links via advanced duplex \cite{yao2020downlink} or non-orthogonal transmission schemes \cite{wu2018spectral}. Another promising solution is to further enhance the resolution of time-frequency resource to facilitate more flexible frame structure and incorporate new modulation schemes, such as index modulation \cite{basar2016index}. With massive combination of frame structures, modulation and duplex schemes, a fast mode adaptation can help to improve the performance as well. As illustrated in Figure~\ref{fig:6g}, each category is able to provide two to three times price reduction.
\end{itemize}

By aggregating different technologies in different domains, we argue that {\bf 1000 times price reduction} is also technically achievable, and the key to success is the AI-assisted intelligent communications, which offer 20 to 50 times reduction in total.

\section{AI-assisted Intelligent Communications} \label{sect:AIIC}
It is a natural question to ask why we can achieve another 20 to 50 times cost reduction using AI-assisted intelligent communications on top of conventional spatial, frequency, and time domain adaptations. A preliminary answer is due to the unique challenges of 6G networks as explained below.

\subsection{PHY: From Stationary to Non-Stationary}
In the traditional wireless system design, an implicit assumption adopted is the stationary variations of physical environments. Taking the famous adaptive modulation and coding schemes (MCS) as an example, the instantaneous adaptation is on the basis of channel conditions, where we rely on the channel reciprocity property in time division duplexing (TDD) systems and the quantized channel feedback in frequency division duplexing (FDD) systems. No matter which scheme is adopted, we assume the obtained channel condition and the wireless fading environment for upcoming transmission follows the similar statistic.

In general, the above stationary assumption is sufficient for relatively low mobility, limited antenna size, and narrow band transmission. As aforementioned, in order to achieve 6G vision, it is necessary to extend the utilization of spatial, frequency, and time domain resources, and the stationary condition no longer hold true. For instance, in the spatial domain, the non-stationary characteristics of channel statistics begin to appear in ultra massive antenna configuration \cite{akyildiz2016realizing}. The similar non-stationary phenomenon happens in the time domain as well \cite{Ghazal2017}, where high mobility is the most important factor based on the current literature. In the frequency domain, in addition to the possible non-stationary effects from the flexible utilization of different duplexing bands, another non-stationary situation may appear when hopping from low frequency bands to high frequency bands including mmWave, THz, and VLC.

The modeling of non-stationary relations among different wireless resources often requires high order non-linear correlation functions, where theoretically optimal approaches are difficult to find in general \cite{yu2018beyond}. With AI-assisted technology, the modeling problem of non-stationary characteristics can be systematically solved via generating sufficient data-sets and machine learning \cite{sugiyama2012machine}. As an example, it has been reported in \cite{eisen2018learning} that the non-stationary channel statistics among different types of base stations can be exploited using deep learning. Although the AI-assisted physical transmission is still in its infancy, a promising throughput gain can be expected by a deep understanding of non-stationary properties of 6G networks.

\subsection{MAC: From Orthogonal to Non-Orthogonal}
Interference is often regarded as the major limitation to improve the network throughput, and one of the major tasks in the MAC layer is to control the inter-cell and intra-cell interference levels through efficient radio resource management. From 1G to 5G networks, the classical MAC strategy relies on the orthogonal utilization of wireless resources to minimize the potential inferences among different links. Thanks to the one-to-one correspondence of radio resources and wireless connections, the associated MAC scheduling complexity is controllable as well.

Nevertheless, the orthogonal strategy requires network entities to accurately predict the dynamics of wireless traffics and prepare suitable wireless resources for the upcoming wireless transmission. This prerequisite holds true in the previous generations, since the most important application scenario is for human-to-human communications and the traffic variations are generally limited by the population density. In order to connect everything with the minimum network cost for 6G, a brute force scheme to turn on massive resources in spatial, frequency, and time domains is definitely impossible and a non-orthogonal management scheme is still necessary to minimize the amount of standby resources. It is also worth mentioning that non-orthogonal wireless transmission has been proposed as a candidate solution for massive grant-free multiple access scenarios and proven to be feasible in cellular fading environments \cite{ding2017application}.

The non-orthogonal usage of wireless resources breaks the one-to-one mapping with radio links, which incurs the {\em Curse of Dimensionality} in designing the optimal multi-user scheduling strategy. Recently, several AI techniques, e.g. deep reinforcement learning \cite{sutton2018reinforcement}, has demonstrated their potential to deal with huge scale optimization problems, such as playing chess or Go games, and the famous AlphaZero can beat top Go player even without the prior human experts' knowledge \cite{silver2018general}. Therefore, it is quite promising to apply AI techniques for solving the complicated non-orthogonal resource management problem as well.

\subsection{NET: From Unified to Customized}
Nowadays, the principle of network layer design for wireless systems is to provide seamless connections to the Internet, and all Internet protocol (IP) based solutions become the inevitable trends. Since the IP based solution is primarily designed for wired networks with limited mobility and negligible packet transmission loss, the current efforts towards all IP networks focus on providing the mobility management capabilities and preventing the wireless network from harmful congestion events \cite{chen2019passive}.

Regardless of the great achievement made in the previous generations, the above unified all IP based framework has the following limitations. First, the negligible packet transmission loss in wired networks is difficult to achieve in wireless scenarios, especially when we consider 6G scenario with diversified fading conditions for massive spatial-frequency-time resources \cite{saad2019vision}. Second, since the quality of service (QoS) guarantee is not the primary goal of the Internet, only a few QoS parameters are supported by the current design, which is definitely insufficient for 6G networks with various quality of experience (QoE) requirements \cite{zhang20196g}. Third, the future network topology will be much more heterogeneous than before. In order to connect everything in 6G, device-to-device, ad hoc, mobile edge/fog intelligence, and other virtualization technologies will be widely deployed \cite{letaief2019roadmap}. Last but not least, there is only limited solution to deal with various security and privacy requirement under the existing Internet environment. With the above reasons, a customized IP based network is thus desired.

The customized design requires more description of the wireless traffics. Due to the limited service quality indicators from the above layers and the data protection regulations from governors, to rebuild a customized flow on top of the existing TCP/IP protocol is not the best choice and an AI based framework to realize the joint efforts of traffic sensing and recognition, QoE predicting, and IP based protocol optimization, will be more suitable than today's unified strategy \cite{cui2014unified}.

\subsection{APP: From Communication to Imagination}
Previous generation wireless networks are mainly in charge of unified information transmission, regardless of the specific running applications. This partially explains why the wireless network operators make less profit than the Web APP runners in the current mobile Internet era \cite{maille2014telecommunication}. In order to provide a sustainable developing environment for 6G networks, to build a more friendly ecosystem between operators and users and explore other possibilities beyond the conventional information exchanging tasks for wireless network operators will be necessary.

It has been predicted in \cite{saad2019vision} that distributed intelligence to enable an autonomous, holographic, and human-machine cooperative society is the key driving application for 6G, which is outside the scope of traditional communications. Take the movement control of cooperative mobile robots as an example, it usually requires an integration of several novel technologies, including objective detection, pattern recognition, as well as seamless high accuracy localization and navigation. In addition to the low latency sharing of movement control instructions among different cooperative robots, high accuracy localization tasks require real time monitoring of indoor wireless signals rather than information exchange with 6G networks. Extensive applications may also cover abnormal event detection for smart cities, such as pedestrian monitoring for emergency management, or remote gesture control via electromagnetic waves.

Although it is impossible to list all the potential applications for 6G networks, one promising trend is to design the service or business oriented network, where AI techniques are the key enabler to identify the service types, to reconfigure the network architecture, and to link the imagination with reality.

\section{Case Studies} \label{sect:CE}
In this section, we provide two case studies to demonstrate the power of AI in optimizing the current wireless transmission technologies as well as new applications.

\subsection{Channel Estimation}
Channel estimation has been considered as a vital component of the modern communication systems \cite{coleri2002channel}. Especially in the future 6G communication networks, THz frequency bands, as a new feature for physical layer transmission, will be adopted to meet the high throughput requirements, which undoubtedly bring new challenges. It is well known that the free space spreading loss increases with frequency, and the THz channel is also characterized by a high molecular absorption, which usually leads to signal distortion and colored noise \cite{guan2016millimeter}. Together with the spreading loss, the molecular absorption results in a greater frequency-selective path loss. Hence, novel channel estimation technologies are required to solve the issues in the THz frequency bands. Actually, the  least square (LS) or minimum mean squared error (MMSE) has shown to be effective and achieve satisfying performance \cite{liu2014channel}. However, the conventional channel estimation interpolation methods \cite{park2017new}, such as linear interpolation (LI), Guassian interpolation (GI) or discrete Fourier transform interpolation (DFTI), cannot fully exploit the associated characteristics and model the non-linear relationship caused by molecular absorption.

\begin{figure}
\centering
\includegraphics[width = 3.4 in]{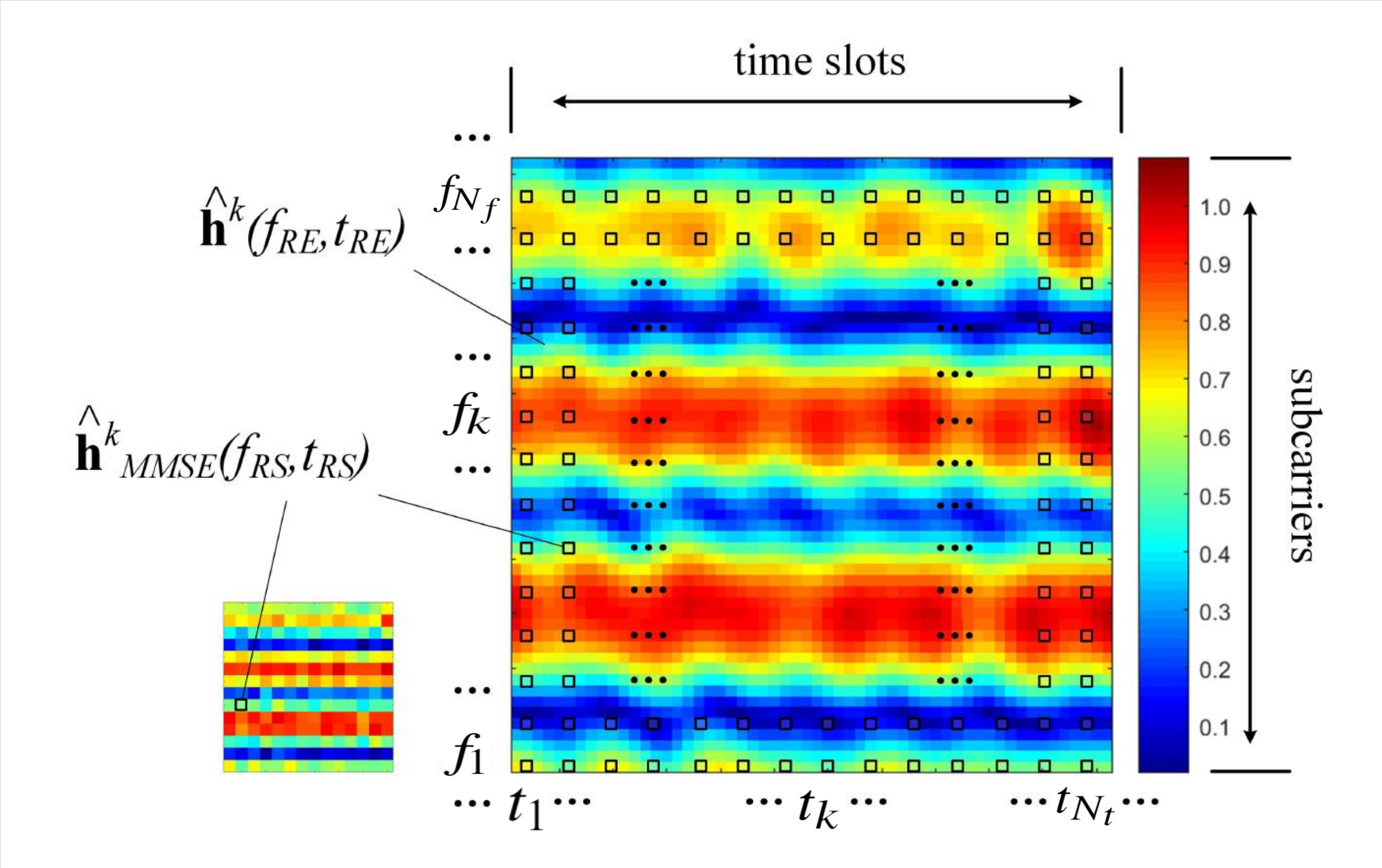}
\caption{The estimated response of one RB with $N_f$ sub-carriers in frequency domain and $N_t$ time slots in time domain. The RE channel state  $\hat{\mathbf{h}}^{k}\left(f_{RE},t_{RE}\right)$ on the right is estimated by the interpolation mechanism from RS channel state $\hat{\mathbf{h}}^{k}_{\text{MMSE}}\left(f_{RS},t_{RS}\right)$ on the left, which is estimated by MMSE algorithms.}
\label{fig:system_ce}
\end{figure}

In order to address this issue, we shall reconsider that our target is to recover the whole channel frequency response from sparse observations at pilot locations. The entire interpolation process is similar to a typical image super-resolution (SR) task \cite{freeman2002example}, which is typically obtaining a high resolution (HR) image from low resolution (LR) images. An interpolation process is usually performed on a resource block (RB) basis with $N_t$ time slots and $N_f$ sub-carriers as shown in Fig.~\ref{fig:system_ce}.
Hence, some existing literature has proposed to borrow SR ideas to solve the interpolation problem in channel estimation processes. For example, traditional image interpolation schemes like bilinear or bicubic, have been utilized to test the performance in \cite{shi2019channel}, which is conducted by assuming linear or non-linear relationships between LR and HR images. To reduce the significant computational complexity of the above algorithms, several learning methods like dictionary learning \cite{yang2010image}, local linear regression \cite{tang2011single} and random forest algorithm \cite{schulter2015fast} manage to directly learn the non-linear interpolation relations of real-time channel state information (CSI) between reference signals and the neighboring resource elements. Although these methods avoid the problem of finding complex and changeable mathematical relationships, the performance in terms of computational complexity and recovering image accuracy, which is measured by normalized mean-square error (NMSE), are not satisfying. To further improve the channel matrix recovery NMSE, deep neural networks based technologies, such as the super-resolution convolutional neural network (SRCNN) \cite{shi2016real}, the very deep super-resolution network (VDSR) \cite{kim2016accurate} and enhanced deep super-resolution (EDSR) \cite{lim2017enhanced} have been applied into channel state interpolation task, which are able to achieve a peak signal-to-noise ratio of more than 30 dB in the conventional image interpolation tasks on the DIV2K public dataset.

\begin{figure}
\centering
\includegraphics[width = 3.4 in]{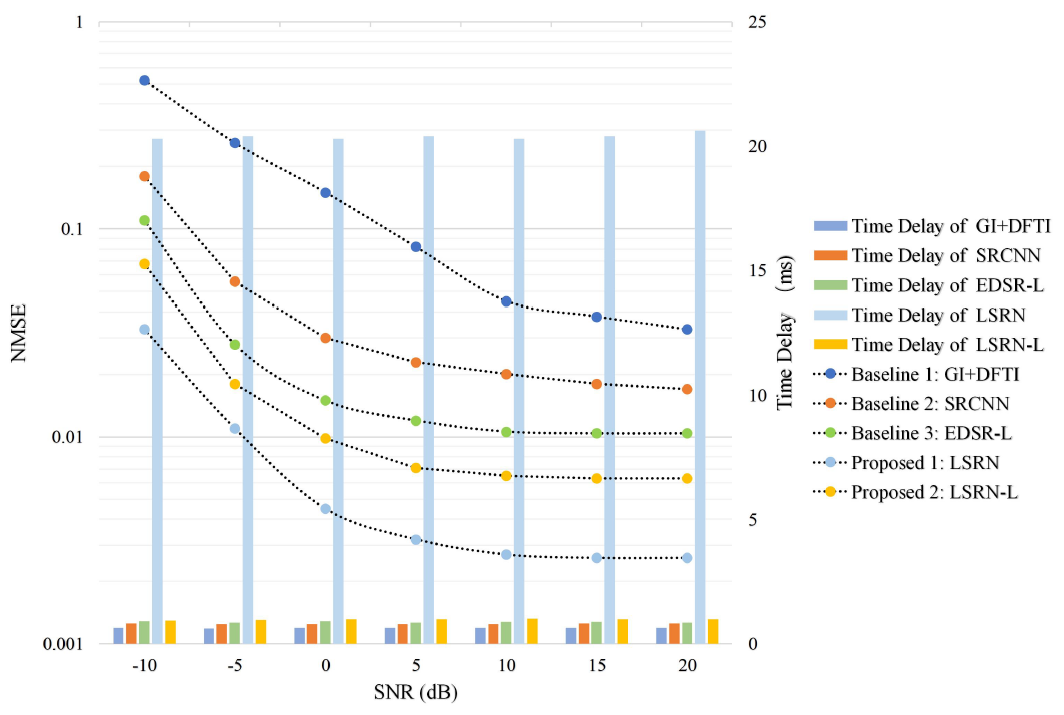}
\caption{NMSE comparison of different interpolation algorithms on the same set of pilot observations, including GI+DFTI, SRCNN, EDSR-L, LSRN and LSRN-L.}
\label{fig:compare_ce}
\end{figure}

The above SR methods are not directly applicable to the practical channel estimation systems, considering the slow-varying properties of channel matrix in the time domain. Hence, the newly proposed schemes utilize a Long-Short term memory based Residual Network (LSRN) with recurrent structure to learn the slow-varying time domain correlation among consecutive OFDM symbols. To make sure the channel estimation performed on a sub-frame (10 milliseconds duration) basis, the modified EDSR and LSRN, named of EDSR-L and LSRN-L with low complexity implementation are proposed, which manage to balance the NMSE performance and the computational complexity. In order to fully compare the performance of different interpolation algorithms mentioned in this case, we perform channel matrix recovery experiments on the same set of pilot observations, using NMSE as the performance indicator. As verified in Fig.~\ref{fig:compare_ce}, the LSRN-L based scheme shows the best NMSE performance over the potential interpolation algorithms with consideration of the acceptable computational complexity and latency requirements.

In conclusion, bring the idea of AI-assisted SR interpolation algorithms brings 20-30 times MMSE reduction for the channel estimation, without affecting the original latency requirements and computational complexity. The noticeable performance improvements in channel matrix recovery make channel estimation a typical case of AI-assisted price reduction in wireless communications.

\subsection{Localization}  \label{sect:LOC}
Localization has already been one of the most popular applications in the modern society. The new proposed 5G communications standard 3GPP Release 16 has mentioned the noticeable high-precision localization vision with NR technology support, which requires that the horizontal positioning error of 80\% of UEs is supposed be less than 3m in indoor deployment scenarios and 10m in outdoor deployment scenarios in commercial use cases \cite{3gppr16}. Predictably, a higher localization accuracy, like decimeter level for regulatory requirements and centimeter level for opportunistic conditions, is required for 6G communications systems, considering the newly adopted features of Thz frequency band and GHz bandwidth. Assuming that such positioning accuracy can be achieved by the new generation communications deployment, the potential applications in the future will be full of imagination. It will bring accurate indoor navigation experience for customers in large shopping malls and exhibition halls, reliable driving assistance information for autonomous vehicles, and seamless tracking for smart manufacturing goods in large factories. However, localization systems with single existing positioning technology are not able to provide such high accuracy as mentioned. The mixture of the mainstream positioning technologies has become a new tendency to improve the positioning accuracy, including 3GPP LTE/NR/6G \cite{zhang2019fingerprint}, WiFi \cite{kotaru2015spotfi,wang2015phasefi,xiang2019robust,wang2017resloc}, Bluetooth Low Energy (BLE) \cite{zhuang2016smartphone}, and Global Navigation Satellite System (GNSS) \cite{marais2005land} technologies. Though practical localization accuracy still depends on the environmental scenarios conditions, multiple sensors fusion can effectively eliminate positioning errors. Fortunately, a large number of smart terminals like smartphones with multiple sensors make it possible for the localization tasks with sensors fusion. Therefore, it is of great importance to review the history of existing technologies and explore the potential of fusion localization.

\begin{figure}
\centering
\includegraphics[width = 3.4 in]{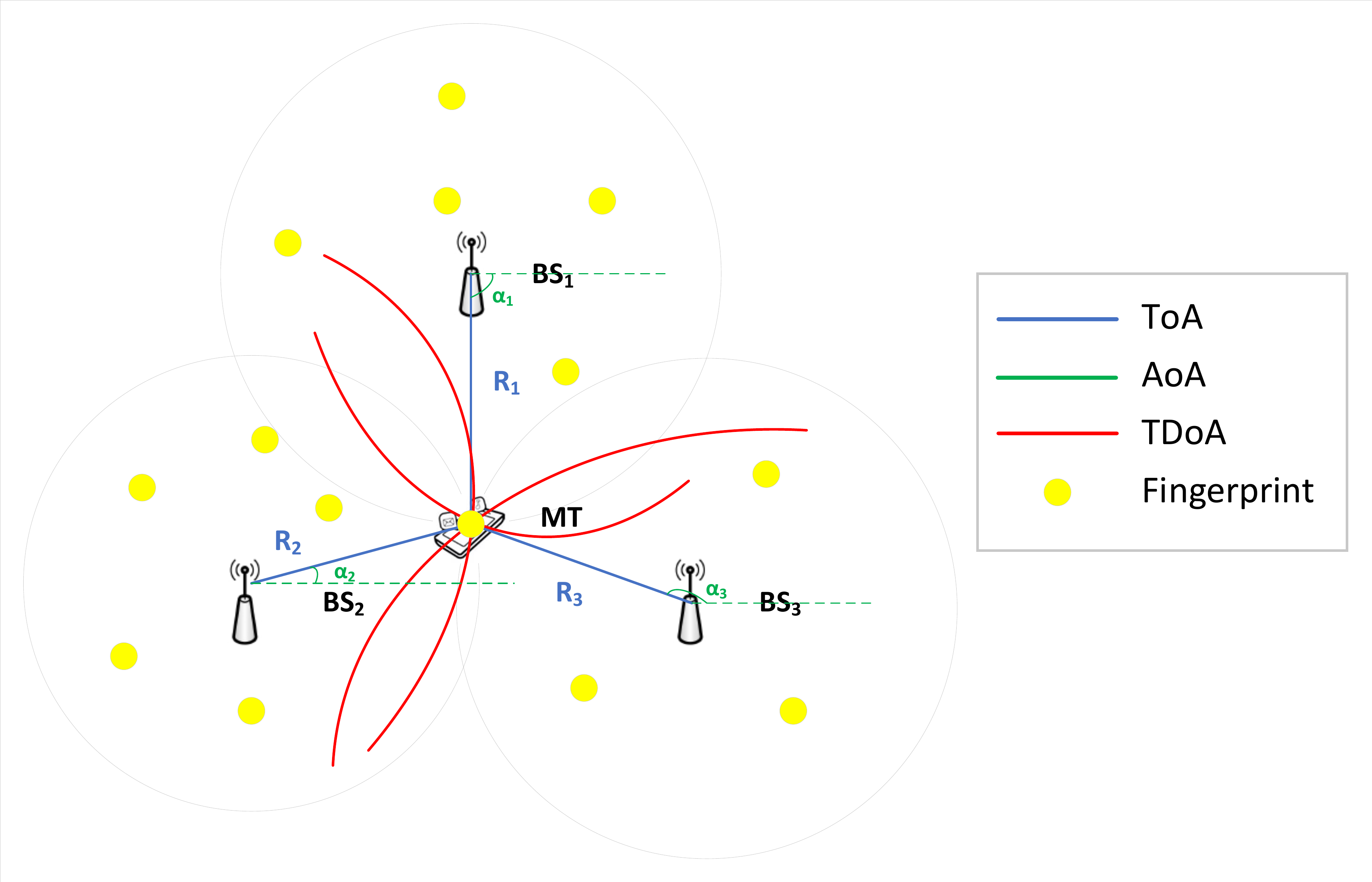}
\caption{ToA, AoA, TDoA and fingerprint based localization methods for mobile terminals are illustrated in blue, green, red and yellow respectively.}
\label{fig:system_local}
\end{figure}

\begin{figure}
\centering
\includegraphics[width = 3.4 in]{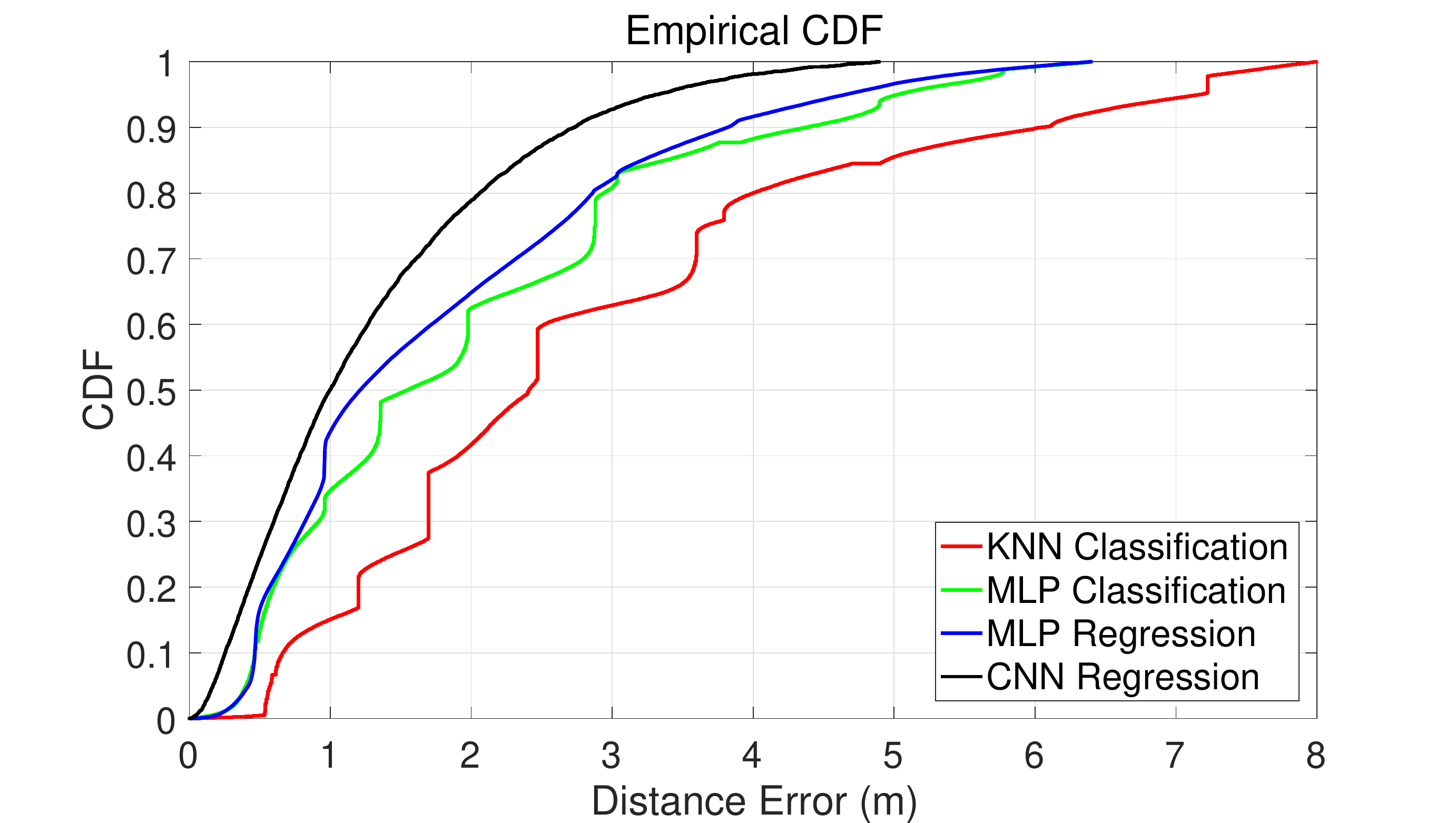}
\caption{CDF of localization errors for different algorithms, including KNN, MLP based classification and MLP, CNN based logistic regression algorithms.}
\label{fig:compare_local}
\end{figure}

Among the existing approaches, triangulation based methods like time of arrival (ToA) \cite{nguyen2016optimal}, time difference of arrival (TDoA) \cite{sun2018solution}, angle of arrival (AoA) \cite{kotaru2015spotfi} are hard to conduct due to the problems of computational complexity and clock synchronization. Conversely, fingerprint based schemes in \cite{paul2009rssi,wang2015phasefi,xiang2019robust,wang2017resloc} have been proven to be an effective solution, where the intrinsic features of wireless signals are extracted in the training stage and utilized in the operating stage to predict the location through real time measured signals. Received signal strength indicator (RSSI) \cite{paul2009rssi}, reference signal received power (RSRP) \cite{zhang2019fingerprint} and CSI \cite{wang2015phasefi,xiang2019robust,wang2017resloc} have been proven to be highly correlated with spatial location and can be utilized as valid location fingerprints. The above localization methods are illustrated in Fig.~\ref{fig:system_local}.
To establishes a probabilistic relationship between the collected fingerprints and the candidate locations, some classifiers such as deterministic k-nearest neighbor (KNN) clustering \cite{xie2016improved} and probabilistic Bayes rule algorithms \cite{wu2017passive} have been adopted, though the corresponding computational overhead during the offline modeling and online feature extraction is usually significant. To address this problem, especially after the deep learning technique has been invented, the model-free localization approaches, such as restricted Boltzmann machine (RBM) \cite{wang2015phasefi}, Multilayer Perceptron (MLP) \cite{tang2015extreme} and convolutional neural networks (CNN) \cite{xiang2019robust} have been applied to exploit fingerprints features and classify to different reference positions with certain probability. According to the latest literature, complex neural network structures has made it to improve the positioning accuracy to the sub-meter, even decimeter level with controllable online prediction overhead. In addition, it is also proven that the similar neural networks, applied to treat the localization tasks as logistic regression tasks rather than classification tasks, can obtain more robust positioning accuracy at non-RP positions \cite{xiang2019robust}. In order to show the performance improvements of AI-assisted localization algorithms more vividly, we summarize the localization accuracy performance results of the existing literature in Fig~\ref{fig:compare_local}. We can find that algorithms with more complex network structures like CNN have better positioning accuracy. Moreover, neural networks applied for logistic regression tasks can effectively eliminate large location errors and show better robustness.

In conclusion, AI-assisted positioning algorithms greatly improve the accuracy performance of fingerprint based localization methods to sub-meter level. Combined with the provided high accuracy, lower computational complexity makes fingerprint based localization stand out from other methods and thus become an important branch of future low-cost localization technologies.

\section{Potential Challenges} \label{sect:cha}
Although it is promising to introduce the AI-assisted technologies for 6G communications, there are still several critical challenges we need to address in the future. In this part, we elaborate them from three key aspects in AI, including data, algorithm, and computing powers.
\begin{itemize}
    \item {\em Wireless Big Data} AI has demonstrated its great success in many computer vision tasks, especially when using supervised learning approach with labelled big data sets, such as ImageNet. In wireless communications, the similar supervised learning method is shown to be effective for solving many complicated optimization problems. However, there are still many concerns on building the public wireless data sets using practical measurement results, even for the research purpose as well.
    \item {\em Portable and Low-latency Algorithm} The existing AI-assisted technologies \cite{jiang2019ai} are designed to satisfy certain performance target, while the migration capability is rarely considered. In order to be widely used in various 6G scenarios, the portability of algorithms should be adopted as an important performance metric. Moreover, due to the urgent delay requirement of many mission critical tasks, the accuracy and latency trade-off of AI algorithms is more desirable than conventional computer vision tasks.
    \item {\em Hardware Co-design} AI-assisted technologies require high density parallel computing resources, which is contradiction with the current implementation methodology of network entities. To facilitate the AI-assist communication, the implementation architecture of wireless networks needs to be reconstructed. In addition, the joint design with the advanced materials such as high temperature superconductor or graphene transistors may also be critical to improve the computing performance.
\end{itemize}

Besides the above listed issues, there are a tremendous amount of other research problems related to AI-assisted communication for 6G networks, such as smart electromagnetic wave control using intelligent surface \cite{huang2019reconfigurable}, or integrated communication, control and intelligent computing design \cite{cao2020unfolded}, which shall bring machine intelligence much closer to human beings. Meanwhile, a fast convergence of quantum communication and computing \cite{mohseni2017commercialize} will provide a powerful platform for AI-assisted technologies. With more convenient human-machine interactions, the 6G communication networks will make themselves a self-reliance and sustainable society rather than cold blooded information pipelines, and we believe the research efforts towards those areas will definitely make a unprecedented success for 6G networks in the near future.

\section{Conclusion} \label{sect:conc}
In this paper, we demonstrate our 6G vision to connect everything by 1000 times price reduction from non-technical customer's viewpoint and discuss the candidate technologies in a well organized manner. In particular, we focus on AI-assisted intelligent communications to illustrate the drive-force behind and the potential challenges. In conclusion, we hope this article can pave the way for identifying the 6G roadmap and aggregate the efforts towards key technology breakthroughs.

\section*{Acknowledgment}
The authors would like to thank Prof. Xin Wang from Fudan University, Prof. Wen Chen from Shanghai Jiaotong University, and anonymous reviewers for their valuable comments, which greatly improves the quality of this paper.

\bibliographystyle{IEEEtran}
\bibliography{IEEEabrv,6G_journal}

\end{document}